\documentclass[a4paper,12pt]{article}
\usepackage{epsfig}
\date{\today}

\newcommand{\bmat}{\left(\begin{array}}
\newcommand{\emat}{\end{array}\right)}
\newcommand{\be}{\begin{equation}}
\newcommand{\ee}{\end{equation}}
\newcommand{\ba}{\begin{eqnarray}}
\newcommand{\ea}{\end{eqnarray}}

\def\lsim{\raise0.3ex\hbox{$\;<$\kern-0.75em\raise-1.1ex\hbox{$\sim\;$}}}
\def\gsim{\raise0.3ex\hbox{$\;>$\kern-0.75em\raise-1.1ex\hbox{$\sim\;$}}}

\def\be{\beta}

\def\lsim{\raise0.3ex\hbox{$\;<$\kern-0.75em\raise-1.1ex\hbox{$\sim\;$}}}
\def\gsim{\raise0.3ex\hbox{$\;>$\kern-0.75em\raise-1.1ex\hbox{$\sim\;$}}}

\begin{document}

\renewcommand{\thefootnote}{\fnsymbol{footnote}}

\rightline{KUNS-1836\textsc{}}
\vskip 1cm
\begin{center}
{\bf \large{  Heterotic string backgrounds and CP violation
\\[10mm]}}
{ Tatsuo Kobayashi $^{1}$ and Oleg Lebedev $^{2}$\\[6mm]}
\small{$^1$Department of Physics, Kyoto University, Kyoto 606-8502,~~Japan\\[2mm]}
\small{$^2$Centre for Theoretical Physics, University of Sussex, Brighton BN1
9QJ,~~UK\\[2mm]}
\end{center}

\hrule
\vskip 0.3cm
\begin{minipage}[h]{14.0cm}
\begin{center}
\small{\bf Abstract}\\[3mm]
\end{center}

In the framework of $Z_N$ orbifolds, we 
discuss effects of heterotic string backgrounds 
 including
discrete Wilson lines 
on the Yukawa matrices and their connection
to CP violation.

\end{minipage}
\vskip 0.7cm
\hrule
\vskip 1cm

{\bf 1. Introduction.}

\vskip 0.5cm

The heterotic string \cite{Gross:1984dd}  
compactified on an orbifold \cite{orbi}
has a number of phenomenologically 
attractive features.
These include the possibility of
obtaining realistic gauge groups and a small number of 
chiral generations at low energies \cite{Ibanez:1986tp},\cite{Ibanez:1987xa}. Also, the Yukawa couplings of the
twisted states exhibit a hierarchy \cite{Dixon:1986qv},\cite{Hamidi:1986vh}   which is a highly
desirable feature from the phenomenological perspective.   

Realistic models require the presence of backgrounds 
\cite{Ibanez:1986tp},\cite{Ibanez:1987xa} --
the Wilson lines \cite{Narain:am} and the antisymmetric background field.
The former are required if we are to obtain a realistic gauge
group and 3 chiral generations, while the latter is 
suggested by  the observed CP violation. 
In this Letter, we study the connection between CP violation and the
backgrounds. In particular, we analyze the effects of the backgrounds
on the Yukawa couplings and whether these effects may lead to 
physical CP violation. We build on earlier work \cite{erler},
\cite{Kobayashi:2003vi}
and now include the effects of discrete Wilson lines and the $U$--moduli.
We also discuss certain subtleties appearing in the 
definition of the CP symmetry from a higher dimensional perspective.

The relevant to the discussion of CP violation 
part of the low energy Lagrangian is given by
\begin{equation}
\Delta{\cal L}= 
Y_{ij}^u H^u Q_{i} U_{j}^c + Y_{ij}^d H^d Q_{i} D_{j}^c \;,
\end{equation}
where $Y_{ij}^{u,d}$ are the Yukawa matrices and $i,j$ are
the generational indices labeling the three chiral families.
Here we exhibit the Yukawa interactions in the two Higgs doublet model,
while for the Standard Model the two doublets are related: 
$H^u \sim (H^d)^c$. When the Higgses develop vacuum expectation
values, these interactions are responsible for generating the quark
mass matrices. These may have complex phases, which can be
absorbed into the definition of the quark fields if
$Y^u_{ij}= \vert Y^u_{ij}\vert e^{i(\alpha_i+\beta^u_j)}$,
$Y^d_{ij}=\vert Y^d_{ij}\vert  e^{i(\alpha_i+\beta^d_j)}$,
but otherwise lead to observable CP violation. 
We will study the Yukawa matrices in $Z_N$ orbifolds which are
defined by dividing a 6D torus by a space group which consists
of the $Z_N$ twists and lattice shifts. We will focus on the 
Yukawa couplings
among twisted states which reside at the orbifold
fixed points $f$
defined by
\begin{equation}
\theta f=f+l\;,
\end{equation}
where $\theta$ is the orbifold twist and $l$ is a torus lattice shift.
The untwisted couplings are moduli independent and do not lead
to CP violation. We will deal mostly with  renormalizable couplings
although some statements about nonrenormalizable couplings 
will be made as well.

The paper is organized as follows. In the next section, we define
the CP symmetry and discuss under what circumstances it is violated
by the backgrounds. In section 3, we present the full moduli dependence of the heterotic Yukawa couplings. Section 4 is devoted to the effects of 
discrete Wilson lines. Section 5 presents our conclusions.

\vskip 1cm

{\bf 2. CP symmetry and the string action.}

\vskip 0.5cm

The bosonic part of the heterotic string action 
in the presence of constant backgrounds
is given 
by\footnote{We omit the pure gauge part of the action which does not pertain
to our considerations.}
\begin{equation}
 S_{\rm }= {1\over 2\pi} \int d\tau d\sigma \Bigl( 
 G_{ij} \partial^\alpha X^i \partial_\alpha X^j -
 B_{ij}  \epsilon^{\alpha\beta} \partial_\alpha X^{ i} 
         \partial_\beta X^{j} +
 A_{ iI} \epsilon^{\alpha\beta} \partial_\alpha X^{ i} 
         \partial_\beta X^{ I}       
\Bigr)\;.
\label{S}
\end{equation}
Here $X^i$, $i$=1,..10 are the space--time coordinates, 
$X^I$, $I$=1,..16 are the gauge space left--moving coordinates, and
$G_{ij}$, $B_{ij}$, and $A_{ iI}$ are the 
background metric, the antisymmetric field, and the Wilson line,
respectively.

In what follows we will discuss CP properties of the twisted Yukawa
couplings which presumably are  the source of the observed CP violation
in the quark sector. CP violation in the Yukawa couplings $Y_{\alpha \beta \gamma}$
is directly related to CP properties of the string action since
\begin{equation}
Y_{\alpha \beta \gamma}= {\rm const}~ \sum_{X_{\rm cl}} e^{-S_{\rm cl}} \;,
\label{string}
\end{equation}
where $X_{\rm cl}$ are solutions to the string equations of motion in the
presence of the twist fields,
$S_{\rm cl}$ is the $Euclidean$ action,
 and $\alpha, \beta, \gamma$ label the twisted
sectors.

Further, we will study a class of the $Z_N$ orbifold compactifications
of the $E_8 \times E_8$ heterotic string
which admit the decomposition $T^2 \oplus T^2 \oplus T^2$
such that the backgrounds have a block--diagonal form.
To discuss the CP symmetry, one introduces the orthogonal
coordinates $X'= O^{-1} X$  in which the metric is diagonal
\begin{equation} 
G ~ \longrightarrow ~ O^T GO = \eta \;, 
\end{equation}
with $\eta $ being the Minkowski metric.
Henceforth, we will work in this basis and will omit the prime.
The bosonic part of the  CP transformation 
can be  defined as
\cite{Strominger:it},\cite{Dine:1992ya}\footnote{See also 
\cite{Lim:1990bp},\cite{Kobayashi:1994ks}.}
\begin{eqnarray}
 X^I &\longrightarrow& -X^I ~, ~ I=1,..16 \;, \nonumber\\
 X^{i}  &\longrightarrow& ~X^{i} ~,~ ~i=1,5,7,9 \;, \nonumber\\
 X^{i}  &\longrightarrow& - X^{i} ~,~ i=2,3,4,6,8,10  \;.
\label{cp}
\end{eqnarray}
This is a combination of the conventional reflection of the three spacial
coordinates, a reflection of three of the compactified coordinates,
and a reversal of the gauge charges.
Although a four dimensional P operation is not a proper Lorentz 
transformation, it becomes one when supplemented by the reflection
of the compactified coordinates. 
In terms of the complex coordinates of the internal manifold,
this amounts to the complex conjugation $Z^i \longrightarrow Z^{i*}$
\cite{Strominger:it}.
The reversal of the gauge charges
is an automorphism of the $E_8 \times E_8$ group and therefore is also a
symmetry of the system (when no backgrounds are present).

The above definition does not appear to be unique in the sense that one can 
extend a conventional P operation to a proper Lorentz transformation
in a number of ways, for instance, through a reflection of only one 
of the compactified coordinates. 
Such ``truncated CP''
appears as a well defined symmetry at the classical 
bosonic  action level, but it is $not$
a symmetry of the theory as a whole. In particular, since it acts
in the 6D subspace, 
it is not a gauge symmetry in the fermionic sector  \cite{Dine:1992ya}.
Also, it is not a symmetry of the compactification:
it transforms the twist $(\theta_1, \theta_2, \theta_3)$ into
 $(\theta_1^*, \theta_2, \theta_3)$ which does not belong
to a subgroup of $SU(3)$ (unless $\theta_1=e^{i\pi}$ ), which
leads to a non--supersymmetric orbifold. Similar arguments
apply to  Calabi--Yau compactifications.

The presence of string backgrounds $B_{ij}$ and $A_{iI}$
$generally$ violates the CP symmetry as defined in Eq.(\ref{cp}).
First, consider the antisymmetric background $B_{ij}$.
Under our factorization assumption, it can be written as
\begin{equation}
 B=B_{(1)} \oplus  B_{(2)} \oplus B_{(3)}\;,
\end{equation}
where $B_{(i)}$ corresponds to the $i$-th compactified plane.
The invariance of the action under twisting 
\begin{equation}
\theta^T B ~\theta =B \;,
\label{B}
\end{equation}
where $\theta$ is the orbifold twist,
and the antisymmetry require
\begin{equation}
B_{(i)}= \left( \matrix{ 0 & {\rm b}_{(i)} \cr
                        - {\rm b}_{(i)} & 0}
\right) \;.
\label{bb}
\end{equation}
Then, clearly if ${\rm b}_{(i)} \not=0$, the symmetry  (\ref{cp})
is violated. On the other hand, this background preserves 
the block--diagonal part of the  (proper) Lorentz symmetry. Indeed,
the orbifold twist splits as
\begin{equation}
 \theta=\theta_{(1)} \oplus  \theta_{(2)} \oplus \theta_{(3)}\;,
\end{equation}
where
\begin{equation}
\theta_{(i)}= \pm \left( \matrix{ \cos\theta_i & \sin\theta_i \cr
                        - \sin\theta_i & \cos\theta_i}
\right)
\end{equation}
rotates the $i$-th plane by $\theta_i$. This is a proper Lorentz
transformation, so Eq.(\ref{B}) signifies the invariance of the B--background
under this class of Lorentz transformations.
However, the reflection symmetry of the background--free action
\begin{equation}
 S_{0 }= {1\over 2\pi} \int d\tau d\sigma 
  ~\partial^\alpha X^i \partial_\alpha X^i
\end{equation}
is lost. Note that this action is invariant under the full Lorentz group
including orientation changing transformations
\begin{equation}
X^i ~ \longrightarrow -X^i
\label{reflection}
\end{equation}
for any $i$.

This observation raises the question ``What is the higher dimensional 
symmetry that ensures that the low--energy 4D Yukawa couplings conserve CP
?''  To answer this question, let us first note that 
4D field theory tells us that complex Yukawa couplings
break conventional CP (if the phases cannot be rotated away)   
while real ones conserve it. From Eq.(\ref{string}) it is
clear that CP is broken when the Euclidean action
has a non--vanishing imaginary part. The 
$\partial^\alpha X^i \partial_\alpha X^i$ piece always gives 
a real contribution, while the $B_{ij}$ contribution is imaginary.
The difference arises from the $\tau$--dependence 
combined with the Wick's rotation:
a quadratic $\partial_\tau$ dependence gives a real result while
a linear one produces a factor of $i$, i.e. we have
\begin{eqnarray}
&&  \partial^\alpha \partial_\alpha  ~~~ {\rm vs}~~~ 
\epsilon^{\alpha\beta} \partial_\alpha \partial_\beta \;.
\nonumber
\end{eqnarray}
The $\epsilon^{\alpha\beta}$--piece leads to antisymmetric
 with respect to the Lorentz (or Lorentz--gauge) indices
contributions to the action.
Such contributions necessarily break some reflection symmetries 
(\ref{reflection}). We thus conclude that the {\it  parity symmetry} 
(\ref{reflection}) ensures that the Yukawa couplings conserve CP.
Under our factorization  $T^2 \oplus T^2 \oplus T^2$ assumption,
this is equivalent to requiring the CP symmetry (\ref{cp}).

Now, it is clear that, from the bosonic action perspective,
the ``truncated CP'' does not reduce to the conventional 4D CP symmetry.
Consider, for instance,  $B_{ij}\not= 0$
in the third plane only. This configuration conserves ``pseudo--CP''
defined with the reflection of one axis
in the first plane only. Yet, the Yukawa
couplings violate CP in the usual sense.

Non--vanishing $B_{ij}$  does not necessarily result
in observable CP violation since at isolated points in the
parameter space the effect of $B_{ij}$ may simply amount
to $S \rightarrow S+ 2\pi i n$. Further constraints come 
from\footnote{Apart from producing CP violating
Yukawa couplings, $B_{ij}$ (with 4D indices) also couples to 
$F_{\mu\nu} \tilde F^{\mu\nu}$ which violates CP. This term is
constrained to be extremely small experimentally leading to the notorious
strong CP problem. }
\\ \ \\
(i) flavor--dependence,
\\ \ \\
(ii) modular invariance.
\\ \ \\
The first of them means that, if we associate quark flavors with
orbifold fixed points,
 the CP violating phases produced by 
$B_{ij}$ should depend on the relative positions of the  fixed points in such a way that they could not be eliminated by a redefinition
of the quark fields \cite{Lebedev:2001qg}.  This amounts to non--vanishing of the
Jarlskog invariant\footnote{In supersymmetric models, CP violation 
is governed by a class of $K$- and $L$- invariants in addition
to the Jarlskog invariant \cite{Lebedev:2002wq}.  } \cite{Jarlskog:1985ht}.
The second requirement (which is related to the first one) means that
the CP phases cannot be eliminated by a modular transformation
whenever the system possesses a modular symmetry \cite{Dent:2001cc},
\cite{Lebedev:2001qg}.
Whether these requirements are satisfied or not depends
on the orbifold and the relevant moduli, but in principle observable CP
violation can be achieved even at the renormalizable level
if $B_{ij} \not=0$ \cite{Lebedev:2001qg}.

The discussion of the Wilson lines proceeds (at first) along similar lines. 
Consider the case of continuous Wilson lines.
A continuous Wilson line is realized through the correspondence
between the space group and rotations and shifts of the gauge
$E_8 \times E_8$ lattice \cite{Ibanez:1987xa}:
\begin{equation}
(\theta, l) ~  \longrightarrow ~ (\Theta,a)  ~~,
\end{equation}
where $\theta$ is a point group element, $\Theta$ is a rotation
of the $E_8 \times E_8$ lattice, and $l$ and $a$ are related by
\begin{equation}
l^{\rm i}=\sum_\alpha n_\alpha {e_\alpha^{\rm i}} \;\;,\;\;
a^{\rm I}=\sum_\alpha n_\alpha A_\alpha^{\rm I} \;.
\label{wl}
\end{equation}
Here $n_\alpha$ are some integers, $e_\alpha^{\rm i}$ are the torus basis vectors,
and $A_\alpha^{\rm I}$ are the Wilson lines.

For the $standard$  embedding ($\theta=\Theta$) , we have 
\begin{equation}
 A=A_{(1)} \oplus  A_{(2)} \oplus A_{(3)}
\end{equation}
where $A_{(i)}$ are $2\times 2$ blocks. 
The twist invariance 
\begin{equation}
\theta^T A ~\theta =A 
\end{equation}
requires
\begin{equation}
A_{(i)}= \left( \matrix{ a_{(i)} & b_{(i)} \cr
                        - b_{(i)} & a_{(i)}}
\right) \;.
\label{ab}
\end{equation}
{\it A priori}, $a_{(i)} \not=0$ or $b_{(i)} \not=0$ violate the symmetry 
(\ref{cp}). However, embedding of the space group into the gauge group
imposes additional constraints.
In particular, on shell,  
effectively we have \cite{Kobayashi:2003vi}
\begin{equation}
\partial_\alpha  X^i_L ~ \sim ~ \partial_\alpha X^I  
\end{equation}
for each of the $2\times 2$ blocks (one can, for instance, take $i$=$I$),
while $X^i_R$ decouple from the Wilson lines.
This identification restores the CP invariance so that  
$S_{\rm cl}$ gives a CP conserving contribution
to the Yukawa couplings. 
These arguments equally apply to a class of non--standard embeddings
for which an orbifold twist is associated with a rotation
of more than one planes in the gauge space.
The explicit formulae  will be given in the next section.

The case of discrete Wilson lines is more complicated and will be dealt
with separately in one of the subsequent sections.

Similar discussion applies to the non--renormalizable couplings.
The relevant $n$--point amplitude is given by \cite{Cvetic:1987qx}
\begin{equation}
A ~ \propto ~ \sum_{X_{\rm cl}} \partial X^i_{\rm cl}...    e^{-S_{\rm cl}}\;. 
\end{equation}
The classical solutions $X^i_{\rm cl}$ are not affected by $B_{ij}$
and $A_{iI}$ since they enter neither the equations of motion nor the boundary conditions. Thus CP violation arises through $e^{-S_{\rm cl}}$
and the arguments above apply.
We note that these couplings are exponentially 
suppressed by the radii of the compactified dimensions in symmetric orbifolds 
\cite{Cvetic:1987qx}.

\vskip 1cm

{\bf 3. Moduli dependence of the  Yukawa couplings.}

\vskip 0.5cm 

To make our arguments more explicit, here
we present the full moduli dependence of the  heterotic Yukawa couplings. 
The Yukawa couplings are calculated via pairing  two real coordinates in 
each plane into a  complex one
(see \cite{Burwick:1990tu} and \cite{erler} for details). 
The action is then written as
\begin{eqnarray}
S_{\rm cl} &=&
{1\over 2\pi} \int d^2 z (\partial Z^i \bar \partial \bar Z^i+
\bar\partial Z^i \partial \bar Z^i ) 
- {B_{i,i+1} \over 2\pi} \int d^2 z (\partial Z^i \bar \partial \bar Z^i-
\bar\partial Z^i \partial \bar Z^i ) \nonumber\\
&+& {1\over 2\pi }\int d^2z \biggl[ 
 {\cal A}_{iI} (   \partial Z^i \bar \partial   Z^I     -
\bar\partial Z^i \partial   Z^I) +
{\cal A}_{iI}' (   \partial Z^i \bar \partial   \bar Z^I        -
\bar\partial Z^i \partial   \bar Z^I) - {\rm h.c.}  \biggr]  \;,
\label{formula}
\end{eqnarray}
where $z=e^{-2(\tau +i\sigma)}$, $Z^i=X^i+iX^{i+1}$, $i=1,3,5\;$; 
$Z^I=X^I+iX^{I+1} \;\;, I=1,3,..,15 \;$; and
\begin{eqnarray}
&&  {\cal A}_{iI}= {1\over 4}(A_{iI}-A_{i+1,I+1}-iA_{i+1,I}-iA_{i,I+1}) 
                                \;,\nonumber\\
&&  {\cal A}_{iI}'= {1\over 4}(A_{iI}+A_{i+1,I+1}-iA_{i+1,I}+iA_{i,I+1}) \;.
\end{eqnarray}
Here h.c. replaces a quantity with the corresponding barred one
and conjugates ${\cal A},{\cal A}'$. We omit the pure gauge
contribution to the action since it vanishes on shell.

The classical solutions are completely determined by their 
singular behavior at the twist operator insertion points and
the boundary conditions. If the twist field of order $k/N$ is 
placed at the point $z_1$ on the world sheet, another
twist field of order $l/N$ is placed at $z_2$, etc., the
solution to the equations of motion  has the form
\begin{eqnarray}
&&\partial Z =c  ( z- z_1)^{-(1-k/N)} 
( z- z_2)^{-(1-l/N)} ( z- z_3)^{-k/N-l/N}\;,\nonumber\\
&&\bar \partial \bar Z =\bar c (\bar z-\bar z_1)^{-(1-k/N)} 
(\bar z-\bar z_2)^{-(1-l/N)} (\bar z-\bar z_3)^{-k/N-l/N}\;,\nonumber\\
&&\bar \partial Z =d (\bar z-\bar z_1)^{-k/N} 
(\bar z-\bar z_2)^{-l/N} (\bar z-\bar z_3)^{-(1-k/N-l/N)} \;,\nonumber\\
&& \partial \bar Z =\bar d ( z- z_1)^{-k/N} 
( z- z_2)^{-l/N} ( z- z_3)^{-(1-k/N-l/N)} 
\end{eqnarray}
for each complex plane. 
The constants $c,d$ are to be determined by the
 the following $monodromy$ conditions
\begin{eqnarray}
\Delta Z^i= \int_{\cal C} dz ~\partial Z^i + \int_{\cal C} d\bar z ~
\bar \partial Z^i= v^i \;, \nonumber\\
\Delta \bar Z^i= \int_{\cal C} dz ~\partial \bar Z^i + \int_{\cal C} d\bar z ~
\bar \partial \bar Z^i= \bar v^i \;,
\end{eqnarray}
with the complex lattice vector $v^i$ defined by $v^i=v^{(i)} +i v^{(i+1)}$
and $\bar v^i=v^{(i)} -i v^{(i+1)}$.
Here the contour ${\cal C}$ is chosen such that $Z^i$ gets shifted but
not rotated upon going around ${\cal C}$.
These equations allow to solve for $c,d$ in terms of the winding
vectors $v^i$.

In the case of the standard embedding, $\bar\partial \bar Z^i$ and 
$\bar\partial\bar Z^I$ have the same 
structure \cite{Kobayashi:2003vi}. 
In particular, they have the same singular behaviour
at the twist operator insertion points:
\begin{equation}
\bar\partial\bar Z^I= \bar c' (\bar z-\bar z_1)^{-(1-k/N)} 
(\bar z-\bar z_2)^{-(1-l/N)} (\bar z-\bar z_3)^{-k/N-l/N}\;,
\end{equation}
while $\partial \bar Z^I= \partial Z^I=  0$ since $X^I$ are left moving.
The constant $c'$ is determined by the monodromy condition for $Z^I$:
\begin{eqnarray}
\int_{\mathcal C} d\bar z~ \bar \partial \bar Z^I = \bar u \;,
\end{eqnarray}
Here $u$ is the gauge space representation of the space group element
$v$ which appears in the monodromy condition for $Z^i$ (see \cite{Kobayashi:2003vi} for 
a detailed discussion).

It has been shown  \cite{Dixon:1986qv},\cite{Hamidi:1986vh} 
that the Yukawa couplings are determined by the holomophic instantons, i.e.
classical solutions with holomorphic $Z^i$ and antiholomorphic $\bar Z^i$
(i.e. $d=0$).
Omitting the intermediate details, let us give the final result \cite{Bailin:nk},\cite{Kobayashi:2003vi}:
\begin{eqnarray}
&&Y_{\alpha\beta\gamma} = {\rm const } \times \label{result} \\ 
&&\sum_{n_i,m_i \in {\rm \bf {Z}} } \exp \biggl[-\sum_{i=1,3,5}
{T_i+ A_i \bar A_i \over  {\rm Re} U_i  }
\Bigl(n_i^2 - 2n_i m_i ~{\rm Im} U_i +m_i^2\Bigr)
{\pi \vert \sin(k_i \pi/N)\vert \vert \sin(l_i\pi/N)\vert \over
  2 \sin^2(k_i l_i \pi/N)  \vert \sin((k_i +l_i)\pi/N)\vert}   \biggr] . \nonumber 
\end{eqnarray}
Here we have used the following definitions of the moduli \cite{Bailin:nk}
\begin{eqnarray}
&& T= {\sqrt{{\rm det}~g }\over 2 \pi^2}   \left( 1 - i   {\rm b}   \right) \;,\nonumber\\
&& U= {1\over g_{11}} \left(  \sqrt{{\rm det}~g }-i g_{12}  \right) \;, \nonumber\\
&& A \bar A = {  \sqrt{{\rm det}~g }  \over 4\pi^2 }    ~ \Bigl\vert a+ib \Bigr\vert^2  
\end{eqnarray} 
for each of the three planes.
Here the integers $n_i,m_i$ are determined by the space group selection rule.
The background parameters b, $a,b$ are given by  Eqs.(\ref{bb}),(\ref{ab}). The 
orbifold metric $g_{ab}$ is 
\begin{equation}
g_{ab}= e_a \cdot e_b
\end{equation}
and we assume  $e_1^2=e_2^2=R^2$ and $e_1 \cdot e_2= R^2 \cos\phi$,
where $R$ is the compactification radius and $\phi$ is an angle 
between $e_1$ and $e_2$.

A comment about the $U$--dependence is in order.
Although we display the $U$--dependence explicitly, in all relevant cases
the value of the $U$--modulus is fixed once the orbifold is specified
($U=-ie^{i\phi}$). 
It is a  continuous modulus only if the 
orbifold possesses a $Z_2$ plane. 
In this case the allowed Yukawa coupling is of the form
$\theta \theta \theta^2$ in this plane, which reduces to a 2--point twist--antitwist
correlator. This gives just a multiplicative constant irrelevant to our discussion. 

It is clear from Eq.(\ref{result}) that the only potential source of CP
violation is the T--moduli (apart from, possibly, discrete Wilson lines which will
be discussed in the next section). 
The U-- and A--moduli only  affect the magnitudes of the Yukawa couplings.
In particular, the presence of continuous Wilson lines always makes the hierarchy
among the Yukawa couplings stronger.

The flavor--dependence necessary for physical CP violation comes from
the dependence of $n_i,m_i$ on the positions of the fixed points
(for analogous discussion of Type I models, see \cite{abel}).
The space group selection rule requires
$n e_1 + m e_2=(1-\theta^{kl})(f_1 - f_2 +\Lambda)$ in each plane,
where $f_{1,2}$ are the fixed points where the fields are placed and
$\Lambda$ is a lattice vector. If the fixed points do not coincide,
$n$ or $m$ do not start from zero and the coupling is suppressed by the 
distance between the fixed points. The consequent CP phases depend
on the relative positions of the fixed points. Given a favorable
configuration,  observable CP violation
can result at the renormalizable level \cite{Lebedev:2001qg}.
This only occurs in the even order orbifolds where the space group selection rule is not
diagonal. In the odd order  orbifolds, CP violation,
if it occurs at the renormalizable level,
 has to result from either
``mixing'' of the fixed points due to an anomalous $U(1)$ or a nonminimal Higgs sector
(e.g. 6 Higgs doublets, etc.). Of course, it can also come entirely from 
non--renormalizable operators in which case not much can be said
quantitatively\footnote{See \cite{nonr}  on related subjects.}.

Concerning the effect of the target space modular symmetries, one can show that, 
due to the axionic shift invariance $T_i \rightarrow T_i +i$,
the CP phases can be rotated away if Im$T_i=\pm 1/2$ \cite{Lebedev:2001qg}.
No CP violation occurs in this case. We note that the axionic shift
symmetry is unbroken by the presence of Wilson lines, unlike the 
duality symmetry, so it is a symmetry of many realistic models.
Thus, under the above assumptions, the T--moduli have to be stabilized away
from the lines Im$T_i=\pm 1/2$ which include the fixed points of the modular group.
This imposes non--trivial constraints on realistic models \cite{Khalil:2001dr}  since the moduli are 
often stabilized at the fixed points $T=1,e^{\pm i\pi/6}$ (as suggested by
symmetries of the scalar potential).

\vskip 1cm

{\bf 4. Discrete Wilson lines and CP violation.}

\vskip 0.5cm 

{\it A priori}, the presence of discrete Wilson lines violates CP.
In this section, we show that this $does$ $not$ occur at least at the renormalizable level
 if the low energy physics is described by the Standard Model (or its minimal
supersymmetric version).

A discrete Wilson line is realized through an abelian embedding 
of the space group into  the gauge group of the orbifold \cite{Ibanez:1986tp}:
\begin{equation}
(\theta,l) ~  \longrightarrow ~ (1, v+a) \;,
\end{equation}
where $v$ is a shift of the $E_8 \times E_8$ lattice, and $l$ and $a$
are related by (\ref{wl}).  If we associate a Wilson line $a_1$ with
$e_1$, then the same Wilson line is also associated with $\theta e_1$.
This can be seen as follows: to respect the group multiplication rules,
one has to associate $(\theta,e_1)(\theta^{-1},e_1)=(1, e_1 + \theta e_1)$
with $(1,v+a_1)(1,-v+a_1)=(1,2a_1)$ from which the above statement follows. 
Thus we have 
\begin{equation}
a^I(e_i) = a^I(\theta e_i), 
\end{equation}
up to a gauge lattice vector. Further, since $(\theta,l)^N=(1,0)$,
such Wilson lines have to be $discrete$:
\begin{equation}
N a^I =0 
\end{equation}
up to a lattice vector.

Consider an example of the $Z_3$ orbifold. 
Denote the two $SU(3)$ root vectors by $e_1$ and $e_2$.
The twist $\theta$ rotates them as follows:
\begin{equation}
\theta e_1 = e_2\;, \qquad \theta e_2 = -e_1 -e_2\;.
\end{equation}
Then the Wilson lines satisfy the following conditions:
\begin{equation}
a_1 = a_2\;, \qquad 3a_1 = 0\;,
\end{equation}
up to a lattice vector. Here we use the shorthand  notation $a_i \equiv a^I(e_i)$.

Similarly, one can study constraints on discrete Wilson lines 
for other orbifolds \cite{Kobayashi:1990mi,Kobayashi:1991rp}. 
The results are presented in Table 1.
The second column shows the orbifold twists, while 
the third column gives  an example of a 6D Lie lattice which
realizes the orbifold twist by its Coxeter element
(in general, there are more than one  6D lattices  realizing the same 
orbifold twist).
The constraints on the discrete Wilson lines depend on 
which 6D lattice is used.
The 6D lattice shown in the third column is the one leading 
 to most degrees of freedom for 
the discrete Wilson lines (see \cite{Kobayashi:1990mi,Kobayashi:1991rp} 
for other lattices).
The constraints are shown in the fourth column, where 
each equation is meant to be satisfied up to a gauge lattice vector.

{\small
\begin{table}
\begin{tabular}{|c|c|c|c|c|} \hline
Orbifold & twist & 6D Lie lattice & Wilson line & further constraints 
\\ \hline
$Z_3$    & $(1,1,-2)/3$ & $SU(3)^3$ & $3a_{1,3,5}
 = 0$ & $a_{i+1} = a_i$, $(i=1,3,5)$ 
\\ \hline
$Z_4$ & $(1,1,-2)/4$ & $SO(5)^2\times SU(2)^2$ & 
$2a_{2,4,5,6} = 0$ & 
$a_1 = a_3=0$ \\ \hline
$Z_6$-I & $(1,1,-2)/6$ & $SU(3)\times G_2^2$ & 
$3a_1 = 0$ & $a_1 = a_2$, $a_{3-6} = 0$ \\ \hline
$Z_6$-II & $(1,2,-3)/6$ & $SU(3)\times SU(2)^2 \times G_2$ & 
$3a_1 = 2a_{3,4}= 0$ & $a_1 = a_2$, $a_{5,6}=0$ \\ \hline 
$Z_7$ & $(1,2,-3)/7$ & $SU(7)$ & 
$7a_1 =0$ & $a_1 =a_{2-6}$  \\ \hline
$Z_8$-I & $(1,2,-3)/8$ & $SO(9) \times SO(5)$ & 
$2a_{4,6} =0$ & $a_{1,2,3,5}=0$  \\ \hline
$Z_8$-II & $(1,3,-4)/8$ & $SO(9)\times SU(2)^2$ & 
$2a_{4-6}=0$ & $a_{1-3}=0$   \\ \hline
$Z_{12}$-I & $(1,4,-5)/12$ & $SU(3) \times F_4$ & 
$3a_1 =0$ & $a_1 = a_2$, $a_{3-6}=0$  \\ \hline
$Z_{12}$-II & $(1,5,-6)/12$ & $SU(2)^2 \times F_4$ & 
$2a_{1,2}=0$ & $a_{3-6}=0$  \\ \hline
\end{tabular}
\caption{Allowed discrete Wilson lines.}
\end{table}
} 

To discuss the discrete Wilson lines further, we will need some facts
about the selection rules for the Yukawa couplings. With every fixed point
let us associate a space group element $(\theta^k,n_i e^i)$ according to
\begin{equation}
\theta^k f_{k,n_ie^i} = f_{k,n_ie^i} + n_ie^i,
\end{equation}
where $n_i$ is an integer. 
Henceforth, we will denote a fixed point by its space group element
$(\theta^k,n_i e^i)$. 
The fixed points $(\theta^p,\ell_ie^i+ (1-\theta^p)\Lambda)$ 
with different $\Lambda$
are equivalent to each other. The twisted states of the orbifold are located 
at the fixed points (in even orbifolds a physical state corresponds to a linear 
combination of the fixed points in the the same twisted sector 
\cite{Kobayashi:1990mc,Kobayashi:1991rp}).

The Yukawa couplings 
among the three states corresponding to 
the fixed points $(\theta^p,\ell_ie^i)$, $(\theta^q,m_ie^i)$ 
and $(\theta^r,n_ie^i)$ are allowed 
only if certain selection rules are satisfied.
In particular, the space group invariance requires that 
\begin{equation}
(\theta^p,\ell_ie^i)(\theta^q,m_ie^i) (\theta^r,n_ie^i)
= (1,\sum_{s=p,q,r}(1-\theta^s)\Lambda) .
\label{selection}
\end{equation}
We note that   
 the right hand side of this 
equation 
is equivalent to $(1,0)$.

As an example, consider this selection rule for 
the $Z_3$ orbifold.
Each 2D $Z_3$ orbifold has three fixed points  
$0$, $e^1/3 +2e^2/3$, and $2e^1/3 +e^2/3$, or in our notation,
$(\theta,0)$, $(\theta, -e^1 - e^2)$, and $(\theta, -e^1)$,
respectively. Since the fixed points shifted by
$(1-\theta)\Lambda= ne^1 +(3m-n)e^2$ are equivalent,
we can denote these fixed points by
$(\theta,k e^1)$ with $k=0,1,2$.
Then  the Yukawa coupling of the states corresponding to 
the fixed points $(\theta,\ell e^1)$, $(\theta,m e^1)$ 
and $(\theta,n e^1)$ is  allowed if 
\begin{equation}
\ell + m + n = 0~{\rm~mod~~3},
\end{equation}
and similarly for the other 2D planes.
This selection rule implies that once $\ell$ and $m$ are specified,
$n$ is fixed uniquely (mod 3). This sort of a selection rule is diagonal
in the sense that the positions of the two of the fixed points determine
the third one uniquely. The resulting quark Yukawa matrices are diagonal.
Note that this orbifold allows for non--trivial Wilson lines (Table 1).

The space group selection rule is not always 
diagonal \cite{Kobayashi:1991rp,Casas:1991ac}. 
Consider the $G_2$ plane 
with the twist $e^{2\pi i/6}$ of the $Z_6$-I orbifold.
Denote the lattice basis vectors as $e_3$ and $e_4$.
The $\theta$ twist has one  fixed point $(\theta,0)$, 
$\theta^2$  has three fixed points 
$(\theta^2,pe_3)$ ($p=0,1,2$), and 
$\theta^3$  has four fixed points 
$(\theta^3,ne_3)$ ($n=0,1,2,3$).
It is easy to show that  the coupling
$(\theta,0)(\theta^2,pe_3)(\theta^3,ne_3)$ 
is allowed for any  $p$ and $n$.
That means that the positions of the two fixed points do not determine the position
of the third one uniquely and the Yukawa matrices can have off--diagonal elements. 
On the other hand, discrete Wilson lines in this plane are not allowed (Table 1).

These two examples suggest that whenever off--diagonal Yukawa matrix elements
are allowed, the discrete Wilson lines are $forbidden$.
This is indeed true as  can be checked by inspecting all orbifolds.
The reason for that is as follows.
A space group element $(\theta^p,n_i e^i)$ 
is embedded into the gauge space as $(1,pv+n_ia^i)$.
The selection rule (\ref{selection}) then implies
\begin{equation}
(1,pv+\ell_i a^i) (1,qv +m_i a^i) (1,rv +n_i a^i)=(1,\Lambda_{E_8 \times E_8})\;.  
\end{equation}
Since $p+q+r=0$ mod $N$ and $Nv$ is a lattice vector, we have
\begin{equation}
(\ell_i +m_i+n_i)a^i=\Lambda_{E_8 \times E_8} \;.
\label{ell}
\end{equation}
Suppose now that $\ell_1=0,1$ are both allowed such that the selection
rule is not diagonal. Taking the difference  of (\ref{ell}) with 
$\ell_1=0$ and $\ell_1=1$, we obtain
\begin{equation}
a^1=\Lambda_{E_8 \times E_8}
\end{equation}
which is equivalent to zero.

This result has important implications. Suppose that the presence of discrete Wilson lines
results in CP phases in the Yukawa couplings. Since in this case the Yukawa matrices
are bound to be diagonal, these CP phases can always be rotated away by a redefinition
of the right handed quarks. Thus, at the renormalizable level, discrete Wilson lines
do not lead to CP violation\footnote{This argument may not apply if there are many
Higgs doublets in the low energy theory (as in Ref.\cite{Abel:2002ih}).}.  

This result may be altered by the presence of non--renormalizable operators
contributing to the Yukawa matrices. 
Although they involve exponentially small factors \cite{Cvetic:1987qx}, 
they are not  necessarily small numerically 
\cite{Cvetic:1998gv} and thus can have a non--negligible effect.
The difficulty with 
an explicit calculation of the  discrete Wilson line  contribution  is that
the standard action (\ref{S}) is not invariant (or does not transform as
$S \rightarrow S+ 2\pi n$) under an $E_8 \times E_8$ lattice shift of the Wilson line
for $arbitrary$ $\partial X^i$. For the same reason, it is not generally twist--invariant. This problem remains open.

\vskip 1cm

{\bf 5. Conclusion.}

\vskip 0.5cm 

We have studied a connection between heterotic string backgrounds and CP violation
in the Yukawa couplings. We find that only the antisymmetric background field $B_{ij}$
is a viable candidate for the source of observed CP violation. The continuous
Wilson lines and the $U$--moduli conserve CP, whereas the discrete Wilson lines
do not lead to physical CP phases at least at the renormalizable level.

{\bf Acknowledgements.}  
T.K. is supported in part by the Grants-in-Aid for 
Scientific Research No.14540256 from the Japan Society 
for the Promotion of Science.
O.L. is supported by PPARC.

\end{document}